\documentstyle[aps,pre,preprint,epsf,epsfig]{revtex}
\begin{document}
\draft
%
%\twocolumn[
%\hsize\textwidth\columnwidth\hsize\csname @twocolumnfalse\endcsname
%

\title{Linear bubble model of abnormal grain growth}

\author{W.W. Mullins}
\address{Department of Materials Science and Engineering,
Carnegie Mellon University, Pittsburgh, Pennsylvania 15213-3890}

\author{Jorge Vi\~nals}
\address{Laboratory of Computational Genomics, Donald Danforth Plant
Science Center, 975 North Warson Road, St. Louis, Missouri 63132}

\date{\today}
\maketitle
\begin{abstract}
A linear bubble model of grain growth is introduced to study
the conditions under which an isolated grain can grow to a size much larger 
than the surrounding matrix average (abnormal growth). We first consider the
case of bubbles of two different types such that the permeability of 
links joining unlike bubbles is larger than that of like bubbles
(a simple model of grain boundary anisotropy).
Stable abnormal growth is found both by mean field analysis and direct
numerical solution. We next study the role of grain boundary pinning 
(e.g., due to impurities or precipitate phases) by introducing a linear 
bubble model that includes lower and upper 
thresholds in the driving force for bubble growth. The link permeability is 
assumed finite for driving forces above the upper threshold, zero below 
the lower threshold, and hysteretic in between. Abnormal growth is also
observed in this case.
\\[0.5cm]
{\em Keywords}: abnormal grain growth, grain boundary anisotropy
\end{abstract}

%\narrowtext
%]

\section{Introduction}

We use linear bubble models of grain growth, originally developed
to study self-similar particle coarsening or the development of texture,
to investigate possible causes of abnormal grain growth. While the
bubble model is a simple idealization of grain growth, it has the 
advantage that the results obtained are not limited by the mean field 
approximation inherent to other existing treatments of abnormal growth.

In normal grain growth
thermal annealing of a polycrystalline material results in self-similar
coarsening driven by excess free energy reduction.
An invariant distribution of scaled grain sizes develops,
with an average grain size that grows as a power law of time
with a characteristic exponent of 1/2 
\cite{re:burke49,re:smith52,re:burke52,re:mullins56,re:mullins98}. 
In abnormal grain growth, on the other 
hand, a few grains grow to a very large size relative to the average matrix. 
In some cases a stable
operating state is achieved characterized by a constant ratio of abnormal
grain sizes to matrix average.

Different mechanisms have been proposed for abnormal grain growth, and
some tested against Monte Carlo simulations of discrete, lattice models
of a polycrystalline material. Defect induced strains can induce isolated
grain growth \cite{re:rollett92}, as well as the same capillary forces
responsible for coarsening when anisotropy of grain boundary energies or 
mobilities exist \cite{re:rollett89,re:grest90}. The conditions for abnormal 
growth due to variable surface energies or mobilities were recently
examined in ref. \cite{re:rollett97} within a mean field treatment of
the matrix grains. For the case of a single grain
with boundary properties that differ from those of the surrounding matrix,
it was found that a higher boundary mobility generally promotes abnormal growth
whereas a higher boundary energy constrains it. The detailed behavior can be
quite complex depending on the ranges of the model parameters chosen.
It includes abnormal growth only up to a limiting grain size, or lower
bounds in the initial size of the grain for abnormal growth to occur.

Abnormal grain growth has also been shown to occur when grain boundaries
pin due to, for example, existing precipitate phases or other defects.
Simplified models have been proposed that introduce grain boundary drag 
forces that lead to ultimate
pinning (Zener pinning) \cite{re:hillert65,re:andersen95}, while the
role of thermal fluctuations to overcome pinning has been analyzed by 
Monte Carlo simulation \cite{re:messina01}. 

Linear bubble models of grain growth were originally introduced by Hunderi 
et al. \cite{re:hunderi78}. Their results showed self-similar coarsening with
parabolic growth kinetics for the average bubble size. The effects of
grain orientation and anisotropic boundary properties have also been 
introduced into these models by Novikov \cite{re:novikov79}, and later 
by Abbruzzese and L\"{u}cke \cite{re:abbruzzese86} to study the development 
of texture. Both in Novikov's work and in later work 
\cite{re:eichelkraut88,re:abbruzzese88,re:mullins93} two types of bubbles A 
and B were considered to represent the idealized situation of only
two different grain orientations. The mobility of
unequal A-B boundaries was assumed to be larger than the mobility of either A-A
or B-B boundaries. The conditions under which a steady state distribution is
reached in this binary case was studied in ref. \cite{re:mullins93}.

We consider in this paper a linear bubble model of grain growth to investigate 
the conditions that could promote abnormal growth in the model. Section 
\ref{sec:anisotropy} presents a mean field
analysis along the lines of the mean field treatment of grain
growth by Rollett and Mullins \cite{re:rollett97}. We consider a
linear chain of bubbles of two types to model grain boundary anisotropy and, 
consistent with their
analysis, show that abnormal bubble growth is possible
when the permeability of links separating unlike bubbles is larger
than that of like bubbles. This mean field analysis is complemented by
a direct numerical solution of the model that confirms the mean field 
predictions regarding abnormal bubble growth: parabolic kinetics 
for both the abnormal bubbles and the matrix, and a constant value of the 
ultimate ratio of abnormal to matrix bubble radii. 

We then explore in Section \ref{sec:hysteresis} a different mechanism that
can lead to abnormal growth even for the case of a single bubble type 
(i.e., in the absence of 
mobility anisotropy). We model grain boundary pinning by introducing
a finite threshold in the driving force for bubble growth. Links between
adjacent bubbles are open if the driving force exceeds an upper 
threshold, and closed if it falls below a lower threshold. In between
the two thresholds we assume hysteretic behavior of the link
permeability. A numerical solution of the model shows
that abnormal growth follows from an initial steady state distribution 
of bubble sizes depending on the values of the upper and lower thresholds.
While the upper threshold
largely determines the subset of bubbles that can grow against the matrix,
we show that there is a sharp transition in behavior depending on
the value of the lower
threshold. Abnormal growth is observed below a critical value, with 
the average radius of the bubbles growing linearly with time. Above this
value, the bubble distribution freezes after an initial transient (growth
stops).

\section{Asymmetric linear bubble model}
\label{sec:anisotropy}

A linear bubble model of abnormal grain growth is introduced
to address the relationship between stable abnormal growth and 
anisotropic grain boundary mobility. The analysis is motivated by
recent research that involved the idealized situation in which a single 
isolated grain $A$ grows in a matrix of $B$ grains \cite{re:rollett97}.
Under the assumptions that the $A-B$ boundary has a different energy and
mobility than $B-B$ boundaries, that the boundary vertices
are in equilibrium, and a mean field treatment of the B grain matrix,
it was concluded that abnormal grain growth is to be observed when unequal
boundaries have higher mobility than equal boundaries, whereas higher
surface energy of unequal boundaries relative to equal boundaries
constrains it. The linear bubble model described here allows us to extend 
these results beyond the mean field approximation for the matrix. We find that
the conclusion that abnormal growth occurs when unequal boundaries have 
higher mobility also holds in this case, 
and that a mean field prediction of the ultimate size ratio is in reasonable
agreement with the results of the numerical calculations.

We consider a set of $N$ spherical bubbles of radii
$R_{i}$, $i = 1, \ldots, N$, forming a linear chain with periodic boundary
conditions. The temporal evolution of the linear bubble model is defined by 
the following set of equations,
\begin{equation}
\label{eq:bubble_def}
\frac{d R_{i}(t)}{dt} = M_{i\;i+1} \left( \frac{1}{R_{i+1}} -
\frac{1}{R_{i}} \right) + M_{i\;i-1} \left( \frac{1}{R_{i-1}} -
\frac{1}{R_{i}} \right),
\end{equation}
where $M_{ij}$ is a permeability coefficient 
between bubbles $i$ and $j$ (the analog of the mobility in 
the grain boundary case). We first consider
in this section the case of two
types of grains, A and B, such that $M_{AA}=M_{BB} = 1$, and define $\mu =
M_{AB}/M_{BB}$. A general property of Eq. (\ref{eq:bubble_def}) is
the existence of a conserved quantity, $\sum_{i=1}^{N} 
R_{i}$, which is independent of time. 

Numerical results for the symmetric case $\mu =1$ were given in ref.
\cite{re:mullins93}. For a random initial distribution of bubble radii, the
ensemble coarsens through growth of bubbles larger than a time dependent
critical radius, and shrinkage and disappearance otherwise.
Following an initial
transient, the configuration reaches a stationary self-similar state. In it,
consecutive configurations of the coarsening structure are geometrically
similar in a statistical sense. As a consequence, any linear scale of the
structure (i.e., the average bubble radius) grows as a power law
of time
$$
\langle R(t) \rangle^{2} - \langle R(t_{0}) \rangle^{2} = C (t-t_{0}),
$$
where $\langle ~ \rangle$ denotes the configuration average, and $t_{0}$ 
is some time in the self-similar regime.

Before presenting the results of our numerical calculations for 
$\mu > 1$, we
discuss a mean field treatment of the linear bubble model (Eq.
(\ref{eq:bubble_def})) along the same lines of ref. \cite{re:rollett97},
and show that similar conclusions follow. We then obtain a numerical
solution of Eq. (\ref{eq:bubble_def}), and demonstrate that, in agreement
with the mean field results, the
linear bubble model does lead to abnormal grain growth when $\mu >1$.
We also show that the ultimate size ratio between the abnormal grains and the
matrix is quite close to that predicted by the mean field analysis.

Consider a bubble of type A in a long chain of B bubbles, and that the AB 
links have a mobility $M_{AB} \neq M_{BB}$, with the mobility ratio 
$\mu = M_{AB}/M_{BB}$. We calculate
the time dependence of $\omega = R_{A}/\langle R_{B} \rangle$ by using a mean
field approximation to the evolution of the B bubbles. We start from,
\begin{equation}
\langle \dot{\omega} | \omega \rangle = 
\frac{d}{dt} \left( \frac{R_{A}}{\langle R_{B} \rangle} \right) =
\frac{1}{\langle R_{B} \rangle^{2}}
\left[ \langle R_{B} \rangle \langle \dot{R}_{A} | R_{A} \rangle - R_{A} \frac{d
\langle R_{B} \rangle }{dt} \right].
\label{eq:apb1}
\end{equation}
Since the A bubble has two B bubbles as neighbors, one has,
\begin{equation}
\langle \dot{R}_{A} | R_{A} \rangle = 2 M_{AB} \left[ \langle
\frac{1}{R_{B}} \rangle - \frac{1}{R_{A}} \right] = 2 M_{AB} \left[
\frac{\alpha}{\langle R_{B} \rangle} - \frac{1}{R_{A}} \right],
\label{eq:apb2}
\end{equation}
where the second equality follows from assuming self-similarity of the 
matrix bubbles and $\alpha =
\langle R_{B} \rangle \langle 1/R_{B} \rangle = \langle R_{B} \rangle /R_{c} 
\simeq 1.1927$ (cf. Appendix \ref{sec:appendixa}). 
The mean field treatment of the B bubbles outlined
in the Appendix gives for the critical radius
$\dot{R}_{c} = M_{BB} /2R_{c}$, and hence from the definition of $\alpha$
we find,
\begin{equation}
\frac{d \langle R_{B} \rangle}{dt} = \frac{M_{BB} \alpha^{2}}{2 \langle
R_{B} \rangle}.
\label{eq:apb5}
\end{equation}

Substituting Eqs. (\ref{eq:apb2}) and (\ref{eq:apb5}) into Eq. (\ref{eq:apb1})
gives,
\begin{equation}
\langle \dot{\omega} | \omega \rangle = \frac{M_{BB}}{\langle R_{B}
\rangle^{2}} G(\mu,\omega),
\end{equation}
where,
\begin{equation}
G(\mu,\omega) = 2 \mu \left( \alpha - \frac{1}{\omega} \right) -
\frac{\alpha^{2} \omega}{2}.
\label{eq:gdef}
\end{equation}
This latter function determines the sign of $\langle \dot{\omega} | \omega 
\rangle$, and therefore whether the A bubble grows or shrinks relative to the 
coarsening B matrix.

\begin{figure}
\psfig{figure=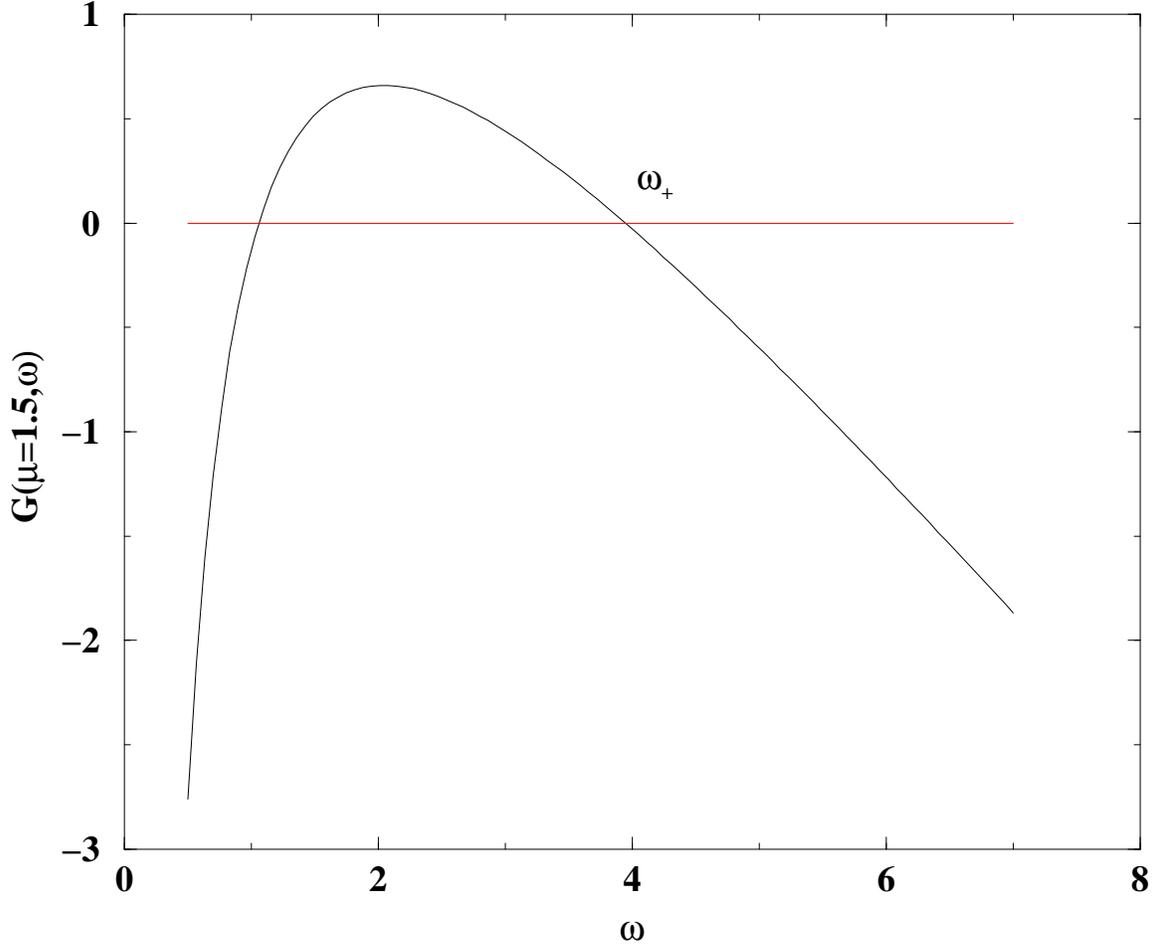,width=6.0in}
\caption{Phase plot $\dot{\omega}$ versus $\omega$ where 
$\omega = R_{A}/\langle R_{B} \rangle$ is the ratio between the radius of
the A and matrix bubbles. $G > 0$ corresponds to ratio growth, and $G < 0$ 
otherwise. The plot shows two fixed points ($G = 0$) at two
different values of $\omega$. The smallest of the two is unstable, and
the largest, denoted by $\omega_{+}$ is stable. This is the expected
operating point of the model and corresponds to a fixed size
ratio between $R_{A}$ and $\langle R_{B} \rangle$.}
\label{fi:omegaphase}
\end{figure}

For $\mu = 1$ the function $G$ is everywhere non-positive. For $\mu > 1$ there
is a range of values of $\omega$ for which $G$ is positive, and in particular
a stable fixed point at some
$\omega = \omega_{+}$ that corresponds to steady abnormal growth.
Figure \ref{fi:omegaphase} shows the phase space plot of $\dot{\omega}$
for $\alpha = 1$ and $\mu = 1.5$. For values of $\omega$ from roughly 1 to
4, $\dot{\omega} > 0$ so that a bubble of type $A$ in this range would
grow relative to the matrix of B bubbles. However, if the ratio $\omega$
exceeds 4, the larger bubble would shrink back to the fixed point. This
root of $G$ is a stable fixed point. The other root 
$\omega \approx 1$ is not stable. The upper root of $G(\mu, \omega)$ is
given by,
\begin{equation}
\label{eq:omegap}
\omega_{+} = \frac{2}{\alpha} \left( \mu + \sqrt{\mu^{2} - \mu} \right).
\end{equation}
The range of relative growth is given by the difference between the upper and
lower roots of Eq. (\ref{eq:gdef}),
\begin{equation}
\Delta \omega = \frac{4}{\alpha} \sqrt{\mu^{2} - \mu}.
\end{equation}

We next compare the results of the mean field calculation to a direct
numerical solution of the set of equations (\ref{eq:bubble_def}). We
only describe the algorithm briefly, further details can be
found in ref. \cite{re:mullins93}. We consider a large number of bubbles
$N = 2 \times 10^{6}$, and impose periodic boundary conditions such that 
$R_{N+1}(t)
= R_{1}(t)$. We initially place 20 equally spaced bubbles of type $A$ in
a matrix of $N-20$ bubbles of type $B$. The initial sizes of B bubbles
are distributed according to the mean field distribution (Eq.
(\ref{eq:pmean_field}) with $\langle R_{B} \rangle (t=0) = 5$). The 
initial radius
of the A bubbles is fixed at $R_{A}(t=0) = 3 \langle R_{B} \rangle
(t=0)$ for the results presented in this section. A wide range of
initial ratios has been investigated with identical results. 
We also set $M_{BB} = 1$. A lower size cut-off $R_{min}$ is introduced 
for numerical reasons so that any bubble
for which $R_{i}(t) \le R_{min}$ during the course of the calculation is
removed, and the two adjacent bubbles redefined as neighbors. The value
of $R_{min} = 0.28$ is chosen so that no bubble can shrink to zero in $\Delta
t = 0.02$, the time discretization used to integrate the system of equations
(\ref{eq:bubble_def}).
The averages shown refer only to averages over the configuration. We have not
performed additional averages over independent initial conditions as the
large number of bubbles considered appears to be sufficient for the
required statistical accuracy.

\begin{figure}
\psfig{figure=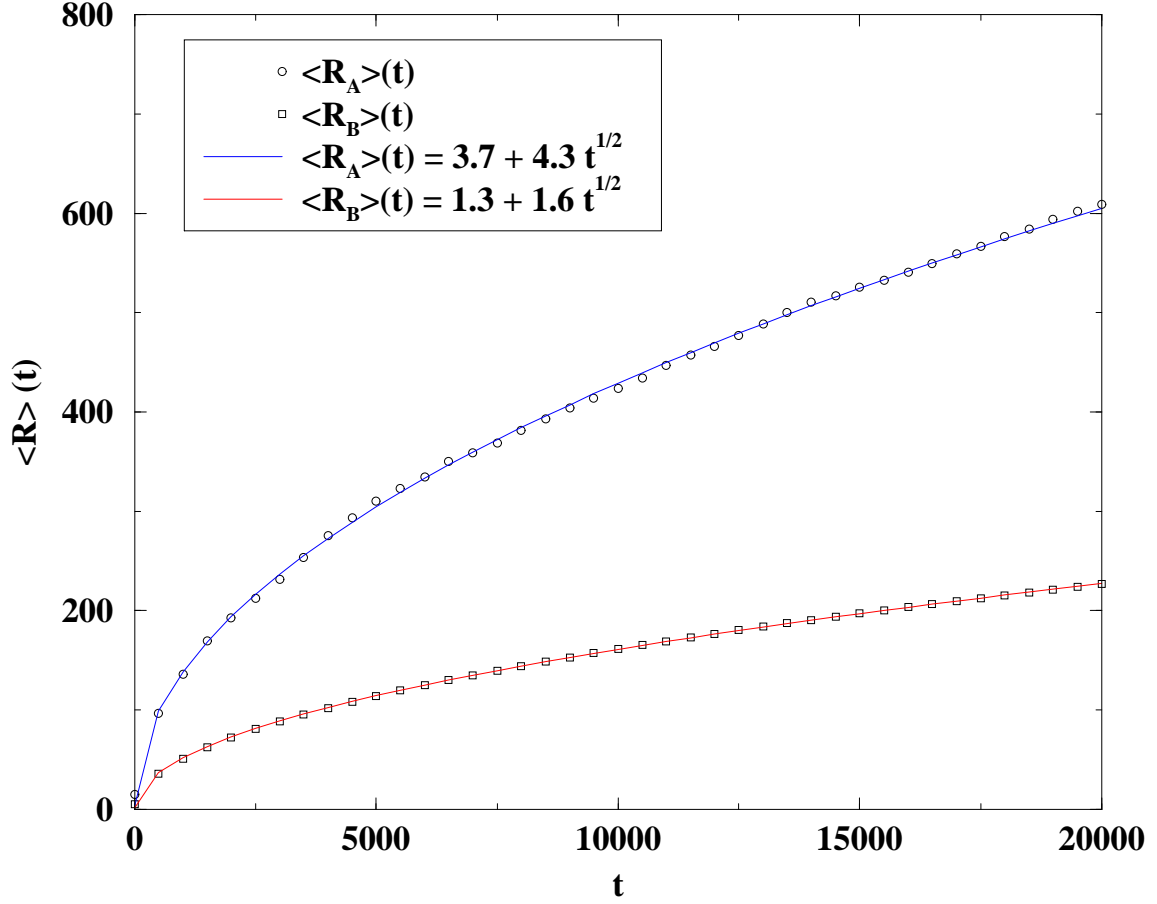,width=6.0in}
\caption{Average radius of large bubbles $\langle R_{A} \rangle$,
and average radius of the matrix $\langle R_{B} \rangle$ as 
a function of time for a
mobility ratio $\mu = 1.5$. Both A and B bubbles exhibit average parabolic
growth to a very good approximation as shown by the fits (solid
lines). The amplitudes of the term $t^{1/2}$ are used to estimate
the quantity $\omega_{+}$ shown in Fig. \protect\ref{fi:omegap} for 
each value of $\mu$.}
\label{fi:mu=1.5}
\end{figure}

\begin{figure}
\psfig{figure=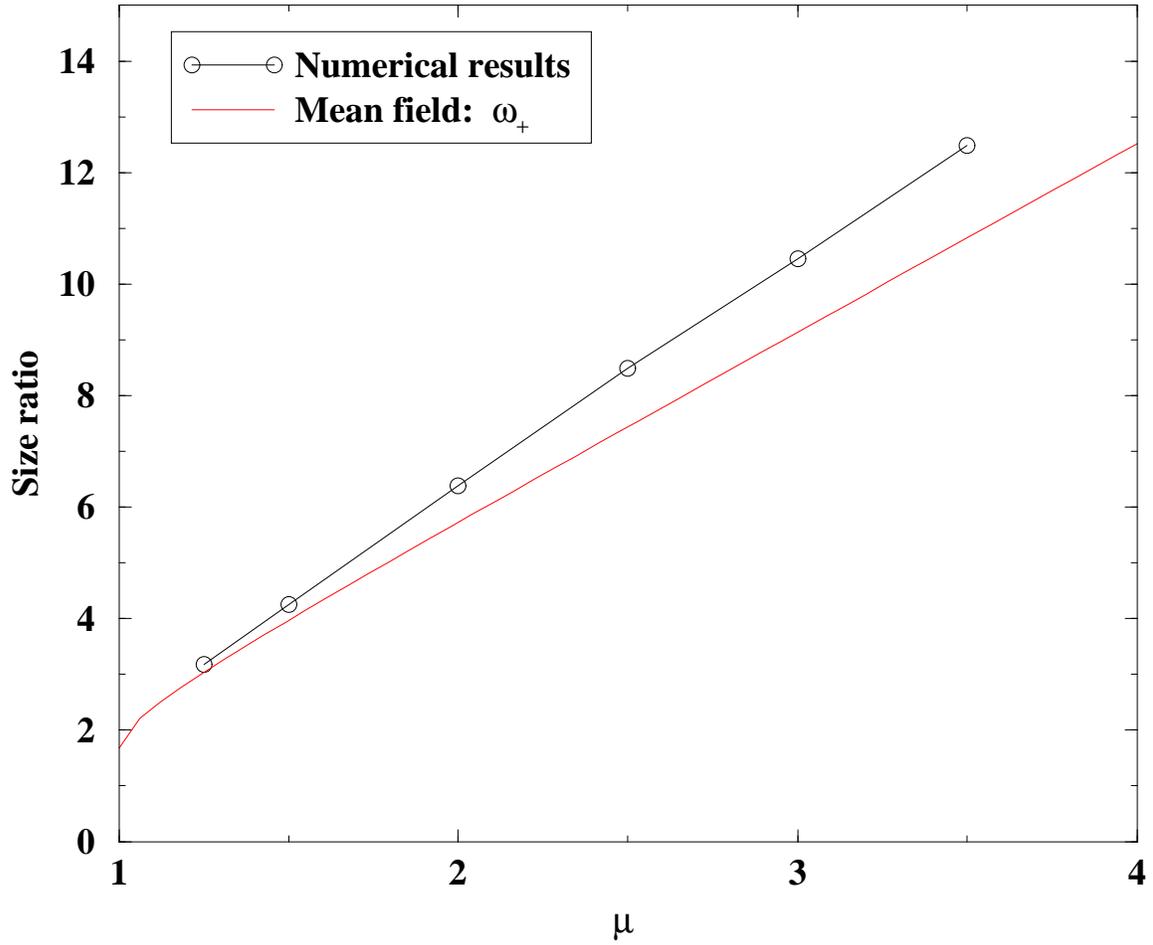,width=6.0in}
\caption{($\circ$), values of $\omega_{+}$ obtained by numerical 
integration. For each value of the mobility ratio $\mu$ the model equations
are integrated in time, and the value of $\omega_{+}$ estimated by
fitting parabolas to the average radii as shown in Fig.
\protect\ref{fi:mu=1.5}. The solid line is the mean field prediction, 
Eq. (\protect\ref{eq:omegap}).}
\label{fi:omegap}
\end{figure}

Figure \ref{fi:mu=1.5} shows our results for $\langle R_{B}(t) \rangle$ 
as well as $\langle R_{A} \rangle $, where the latter is an average over
the 20 bubbles of type $A$. The average radius of the
matrix bubbles $\langle R_{B}(t) \rangle$ exhibits 
power law growth with an exponent of 1/2, in agreement with the mean
field prediction of Appendix \ref{sec:appendixa}. The figure also shows 
two least square fits to obtain
the corresponding amplitudes of the power laws which are used to calculate the
ratio $\langle R_{A} \rangle / \langle R_{B} \rangle$ at long times.
Figure \ref{fi:omegap} shows our numerical results for the ratio
of amplitudes for a range of
mobility ratios $\mu$, and compares them with the mean field prediction
given by Eq. (\ref{eq:omegap}).

In summary, a mean field treatment of the linear bubble model with unequal 
boundary mobilities predicts that abnormal bubble growth will occur for 
$\mu > 1$ with an ultimate size ratio of $\omega_{+}$.  
The numerical results confirm power law growth in time of both A and B average
radii, with an exponent of 1/2. The numerical results for the ultimate
size ratio $\omega_{+}$ are also in excellent agreement with the mean field
prediction. Clearly, bubble size correlations that
are not taken into account in the mean field treatment must only introduce
small corrections.

\section{Symmetric case with a mobility threshold}
\label{sec:hysteresis}

We investigate in this section a different mechanism leading to
abnormal grain growth even in the absence of any mobility anisotropy.
We hypothesize that if a finite threshold to grain boundary motion
exists, then it is possible that a large fraction of the matrix grains
would remain immobile, except for those that were sufficiently larger than 
their neighbors so that the local driving force for growth exceeds the given 
threshold. The excess energy that is contained in
the initial particle distribution would then be relieved mostly through
size increases of the larger grains at the expense of a largely
immobile, high energy, matrix distribution.

In order to investigate this possibility within the linear bubble
model introduced in Section \ref{sec:anisotropy}, we consider an 
ensemble of like bubbles and introduce two threshold values for
the mobility $M$ in Eq. (\ref{eq:bubble_def}). Let $\Delta p = 1/R_{i} - 
1 /R_{i+1}$ be the local driving force associated with the $i-th$ link,
and $\Delta p_{l} < \Delta
p_{u}$ the low and high driving force thresholds respectively. We define
$M = 1$ if $|\Delta p| > \Delta p_{u}$, and $M = 0$ if $|\Delta p| < 
\Delta p_{l}$. We also assume a hysteresis loop in $\Delta p_{l} <
| \Delta p | < \Delta p_{u}$ with $M=0$ in the lower branch and $M=1$ in the
upper branch. Therefore a link remains closed ($M = 0$) until $|\Delta p|$ 
across
the link exceeds $\Delta p_{u}$. Once the link is open ($M = 1$) it remains
open until $|\Delta p|$ falls below $\Delta p_{l}$. Finally, when a
bubble radius falls below $R_{min}$, so that the bubble is removed from
the distribution, a new link between the new neighboring bubbles is
made, and its mobility is assigned to be 1 unless $|\Delta p| < \Delta
p_{l}$.

We have used the same numerical algorithm described in Section 
\ref{sec:anisotropy} to integrate the system of equations 
(\ref{eq:bubble_def}) with the mobility thresholds just introduced.
In this Section we consider an ensemble of
$N = 10^{6}$ identical bubbles, initially distributed according
to the mean field result, Eq. (\ref{eq:pmean_field}), with
$\langle R(t=0) \rangle = 5$. Although
all bubbles are identical in the present case, it is convenient for the sake
of the discussion to refer to those
that grow relative to the average as A bubbles, and as B or matrix bubbles 
to the rest.

\begin{figure}
\psfig{figure=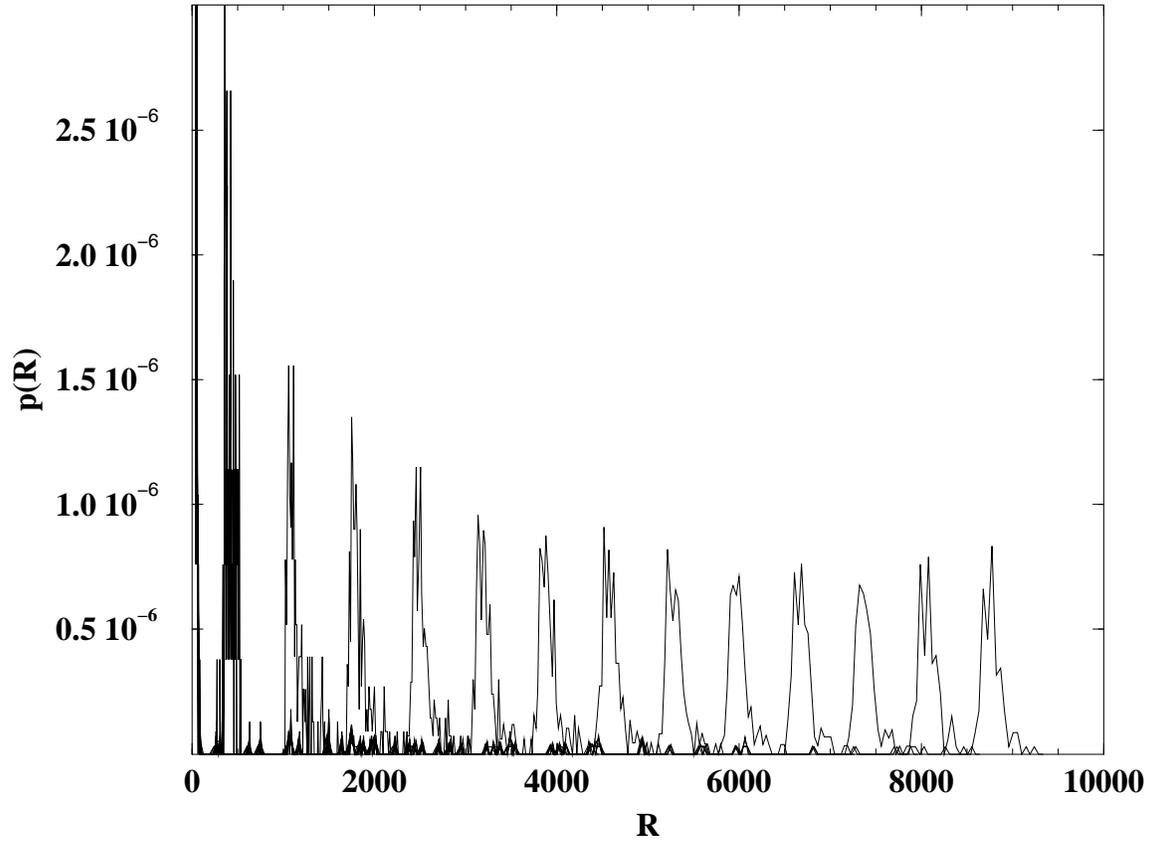,width=6.0in}
\caption{Probability distribution function of particle radius as a function
of $R$ for several times ranging from $t = 1000$ to
$t = 25000$ in increments of 2000 time units. The main peak of the 
distribution has been removed for clarity. Each of the peaks shown at large 
$R$ corresponds to a specific time, and they are ordered from left to right 
according to increasing times. The distributions show the existence of
a small set of large bubbles that grow to a size much larger than the
matrix average.}
\label{fi:p50}
\end{figure}

Figure \ref{fi:p50} shows our results for the bubble radius distribution 
function
for a representative set of parameters $\Delta p_{l} = 0.05$ and 
$\Delta p_{u} = 3.0$. Even though all bubbles are identical and 
follow the mean field distribution at $t=0$, the largest bubbles in the
initial
ensemble grow while most of the rest remain stagnant. This figure shows
the radius distribution $p(R)$ (with the main peak near $R = \langle R
\rangle$ suppressed for clarity) starting at $t = 1000$ all the way up
to $t = 25000$ in increments of 2000 time units. It is clear from the 
figure that a small subset of the initial distribution grows as indicated
by the successive peaks of $p(R)$ at large $R$.

Not all possible combinations of $\Delta p_{l}$ and $\Delta p_{u}$ result
in abnormal growth however. First, there is an obvious upper bound for
$\Delta p_{u}$ given the initial radius distribution, and it corresponds to
the driving force $\Delta p$ between the largest possible bubble 
$R_{max} \simeq 8.4$ for our initial distribution, and $R_{min} \simeq 0.28$.
We find
$\Delta p_{max} = 1/R_{min} - 1/R_{max} \simeq 3.42$. If $\Delta p_{u} > 
\Delta p_{max}$ no bubble will grow.

\begin{figure}
\psfig{figure=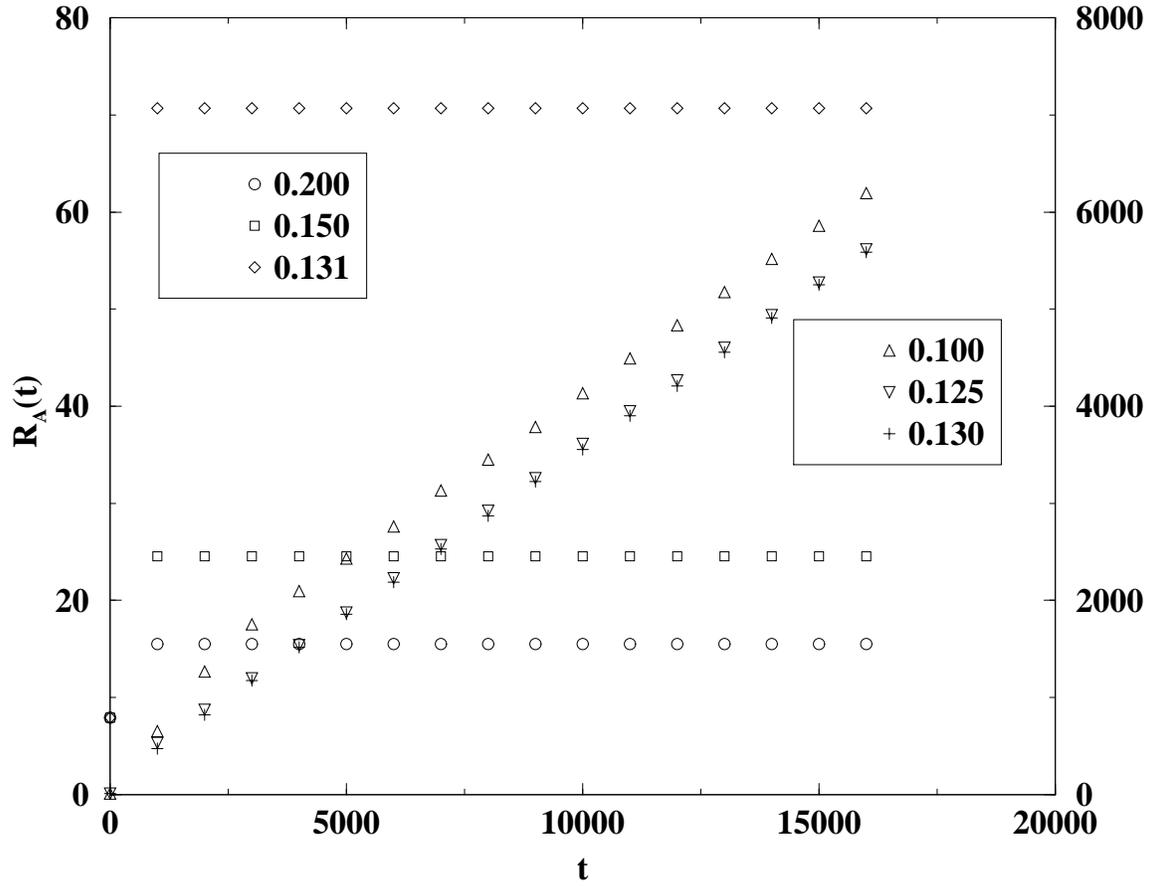,width=6.0in}
\caption{Largest bubble radius in the distribution as a function of time.
The upper threshold $\Delta p_{u} = 2.0$ in all the cases shown, and
the values of the lower thresholds are indicated in the figure. The three 
largest values of $\Delta p_{l} = 0.131, 0.150$ and 0.200 lead to an 
asymptotically stagnant configuration (left axis), whereas the values
$\Delta p_{l} = 0.100, 0.125$ and 0.130 lead to abnormal growth (right axis).}
\label{fi:lanal}
\end{figure}

For fixed $\Delta p_{u} < \Delta p_{max}$, a bubble will grow (call it A)
at the expense of a B neighbor when $R_{A} > \left( 1/R_{B} - \Delta 
p_{u} \right)^{-1}$. Therefore for a given initial distribution the
value of $\Delta p_{u}$ determines the range of radii of bubbles expected to
grow. Once a given A bubble starts growing, it will only stop if it
encounters a bubble B such that $1/R_{B} - 1/R_{A} < \Delta p_{l}$.
Equivalently, 
whenever an A bubble encounters a bubble of radius $R_{B} = \left(
1/R_{A} + \Delta p_{l} \right)^{-1}$ or larger, growth will stop. If one
further assumes that growth of A has already occurred for some time
so that $R_{A}$ is sufficiently large, then
this condition is approximately $R_{B} \simeq 1/\Delta p_{l}$ independent
of $R_{A}$, relation that can be used to defined a critical value for growth 
$\Delta p_{l}^{c}$. If $\Delta p_{l} \ge \Delta p_{l}^{c}$ 
there is a 
nonzero probability that a large and growing A bubble will become 
the neighbor of a
B bubble that is sufficiently large to stop growth of the A bubble. 
If the matrix has remained approximately stagnant, this critical value can
be obtained from $R_{max} \simeq 8.4$ as $\Delta p_{l}^{c} = 1/R_{max} 
\simeq 0.12$
(in practice, the numerically sampled initial condition typically
has $R_{max} \simeq 7.9$ or $\Delta p_{l}^{c} \simeq 0.127$). 
This behavior is observed numerically and 
is illustrated in Fig. \ref{fi:lanal}. The figure shows the radius of the 
largest bubble in the ensemble as a function of time. The upper cut-off 
in all the cases shown is
fixed $\Delta p_{u} = 2.0$, and the figure shows the results for a range
of values of $\Delta p_{l}$. The value $\Delta p_{l} \simeq 
0.130$ marks the transition between abnormal growth and an
ultimately frozen 
configuration. Note that the transition is quite sharp as a function of
$\Delta p_{l}$. Identical numerical results concerning this transition 
as well as the same critical value $\Delta p_{l}^{c}$ have been 
obtained for $\Delta p_{u} = 2.5$ and $\Delta p_{u} = 3.0$.

In summary, the upper cut-off $\Delta p_{u}$ determines the fraction
of the ensemble that can grow, and therefore the degree of stagnation of the
matrix. Once abnormal grain growth has started, the value of the lower
cut-off $\Delta p_{l}$ (and the amount of growth in the matrix, if any) 
determines whether abnormal growth continues or rather the system reaches
a frozen configuration.

We finally mention that while abnormal growth occurs, the typical radius 
$R_{A}$ of the large particles is expected to grow linearly with time. In 
mean field, a given A bubble will have two B bubbles as nearest neighbors and 
therefore
\begin{equation}
\label{eq:ramean}
\frac{dR_{A}}{dt} = 2 M_{AB} \left( \frac{1}{R_{B}} - \frac{1}{R_{A}} \right).
\end{equation}
While bubble A grows successive B neighbors will shrink to zero and be 
eliminated from the ensemble. Therefore the growth of A can
be estimated by averaging Eq. (\ref{eq:ramean}) over the distribution of
B, and when the matrix is almost stagnant, over the initial distribution
of bubble radii. In either case, $\langle 1/R_{B} \rangle$, 
where $\langle ~ \rangle$ denotes average over the configuration, will
be constant (or changing very slowly compared with the rate of growth
of the A bubble), so that for sufficiently long times ($1/R_{A} \ll
\langle 1/R_{B} \rangle$) $d R_{A}(t) / dt$ is approximately constant. 
This is the dependence shown in Fig. (\ref{fi:lanal}) in the cases in which 
abnormal grain growth is observed. The value of the slope of the line 
$R_{A}(t)$ versus $t$ can be straightforwardly obtained by computing the
average $\langle M_{AB}/R_{B} \rangle$ over the initial distribution of 
matrix grains.

\section*{Acknowledgments}
This research has been supported by the U.S. Department of Energy, contract
No. DE-FG05-95ER14566.

\bibliographystyle{prsty}
\bibliography{$HOME/mss/references}

\newpage
\appendix
\section{Mean field calculation for a linear bubble model}
\label{sec:appendixa}

Consider a set of spherical bubbles arranged along a line that model a
set of neighboring (identical) grain boundaries. The equation of motion
for their radii is,
\begin{equation}
\frac{d R_{i}}{dt} = M \left( \frac{1}{R_{i+1}} +
\frac{1}{R_{i-1}} - \frac{2}{R_{i}} \right).
\label{eq:main_eq}
\end{equation}
The mean field approximation for the average (ignoring correlations) is,
\begin{equation}
\langle \dot{R}_{i} | R_{i} \rangle = 2 M \left( \frac{1}{R_{c}}
- \frac{1}{R_{i}} \right),
\label{eq:rdot1}
\end{equation}
where one defines a critical radius through $1/R_{c} = \langle 1/R
\rangle$. Define now a reduced radius $r=R/R_{c}$. Equation (\ref{eq:rdot1}) can
be written as,
\begin{equation}
R_{c} \langle \dot{R}_{i} | R_{i} \rangle = 2 M \left( 1 - \frac{1}{r}
\right).
\label{eq:rdot2}
\end{equation}
As is standard in the analysis of steady state solutions for the averages
\cite{re:mullins91}, one first defines the quantity,
$$
y = \dot{R}_{c} R_{c} = \frac{1}{2} \frac{dR_{c}^{2}}{dt},
$$
so that Eq. (\ref{eq:rdot2}) can be written as,
\begin{equation}
R_{c}^{2} \langle \dot{r} | r \rangle = f(r) - ry,
\label{eq:rdot3}
\end{equation}
where we have defined $f(r) = 2M (1 - 1/r)$. This equation is a
particular case of Eq. (7) in ref. \cite{re:mullins91}. The nodal curve defined
by $\langle \dot{r} | r \rangle = 0$ is thus given in our case by,
\begin{equation}
y = 2 M \left( \frac{1}{r} - \frac{1}{r^{2}} \right).
\label{eq:nodal}
\end{equation}
According to the classical mean field treatment of Lifshitz and Slyozov
\cite{re:lifshitz61}, there exists a stable operating point of the reduced
particle size distribution determined by Eq. (\ref{eq:main_eq}) that 
corresponds
to the maximum of the nodal curve $y=y_{m}$, so that the distribution of 
reduced radii $r$ extends from $r=0$ to a sharp cutoff $r=r_{m}$. For our
particular form of the nodal curve, Eq. (\ref{eq:nodal}), we have $r_{m} = 2$,
and $y_{m} = M / 2$. A statistical self-similar distribution is reached
with this value of $y$, and from its definition, we have,
$$
\frac{dR_{c}}{dt} = \frac{M}{2R_{c}},
$$ that after integration leads to the asymptotic parabolic growth law,
$$
R_{c}^{2}(t) - R_{c}^{2}(t_{0}) = \frac{M}{2} (t - t_{0}),
$$
where $t_{0}$ is some time within the self-similar regime.

The distribution of reduced particle sizes can also be computed by using our
result for $f(r)$, and Eq. (14) in \cite{re:mullins91}. Define the function
$F(r) = t \langle \dot{r} | r \rangle$, which satisfies in the steady state,
\begin{equation}
F(r) = \frac{f(r) - r y_{m}}{2 y_{m}} = - \frac{(2-r)^{2}}{2r}.
\label{eq:fr}
\end{equation}
The general solution of the continuity equation for $n(r,t)$, the number
particle density, is given by \cite{re:mullins91},
$$
n(r,t) = \frac{1}{F(r)} \Psi \left[ t - \Theta(r) \right],
$$
where $\Psi$ is an arbitrary function and where,
\begin{equation}
\Theta(r) = \int_{0}^{r} \frac{dr'}{F(r')}.
\label{eq:theta}
\end{equation}
Substitution of Eq. (\ref{eq:fr}) into Eq. (\ref{eq:theta}) yields,
\begin{equation}
\Theta(r)/2 = - \ln (2-r) - \frac{2}{2-r} + \ln 2 + 1.
\label{eq:theta2}
\end{equation}
With this result, the normalized probability distribution function 
$P(r) = n(r,t)/ \int n(r,t) dr$ is time independent and given by (Eq. (20) in
ref. \cite{re:mullins91}),
\begin{equation}
P(r) = \frac{2e \; r}{(2-r)^{3}} e^{- 2/(2-r)}, \quad 0 \le r \le r_{m}.
\label{eq:pmean_field}
\end{equation}
We note that the upper cut-off is $r_{m}=2$, that the maximum of $P(r)$ occurs
at $r = \sqrt{2}$, and that the average reduced radius is given by $ \langle r
\rangle = \langle R \rangle / R_{c} \simeq 1.1927$.

\newpage

\newpage

\newpage

\newpage

\newpage

\end{document}